\documentclass[aps, twocolumn, superscriptaddress]{revtex4} 

\usepackage[switch]{lineno}
\usepackage{amsmath}
\usepackage{amssymb}
\usepackage{empheq}
\usepackage[parfill]{parskip}
\usepackage[active]{srcltx}
\usepackage{color}
\usepackage{array}
\usepackage{booktabs}
\usepackage{amsfonts}
\usepackage{dsfont}
\usepackage{graphicx}

\begin{document}

\setlength{\parindent}{0.5cm}

%\title{Solvable model of driven matter with pinning}

\title{Solvable model of driven matter with pinning}

\author{Gourab Kumar Sar}
\email{mr.gksar@gmail.com}
\affiliation{Physics and Applied Mathematics Unit, Indian Statistical Institute, 203 B. T. Road, Kolkata 700108, India}

\author{Dibakar Ghosh}
\email{diba.ghosh@gmail.com}
\affiliation{Physics and Applied Mathematics Unit, Indian Statistical Institute, 203 B. T. Road, Kolkata 700108, India} 

\author{Kevin O'Keeffe}
\email{kevin.p.okeeffe@gmail.com}
\affiliation{Senseable City Lab, Massachusetts Institute of Technology, Cambridge, MA, USA, 02139}

\begin{abstract}
\hspace{1 cm}  (Received XX MONTH XX; accepted XX MONTH XX; published XX MONTH XX) \\ 

We present a simple model of driven matter in a 1D medium with pinning impurities, applicable to magnetic domains walls, confined colloids, and other systems. We find rich dynamics, including hysteresis, reentrance, quasiperiodicity, and two distinct routes to chaos. In contrast to other minimal models of driven matter, the model is solvable: we derive the full phase diagram for small $N$, and for large $N$, derive expressions for order parameters and several bifurcation curves. The model is also realistic. Its collective states match those seen in the experiments of magnetic domain walls, and its force-velocity curve imitates those of superconductor vortices.
\\
\noindent \\
DOI: XXXXXXX
\end{abstract}

\maketitle
%%%%%%%%%%%%%%%%%%%%%%%%%%%%%%%%%%%%%
%\section{Introduction}
Driving matter through disordered environments has diverse applications in science. Magnetic domain walls and other quasi-particles may be driven off material defects and used as memory units in spintronics \cite{allwood2005magnetic,luo2020current,wiesendanger2016nanoscale,reichhardt2022statics}. Electromagnetic colloids may be forced to self-assemble into cargo carriers for high precision medicine \cite{reddy2012magnetic,yang2017autocatalytic,bricard2015emergent,caleap2014acoustically}. The colloids may also be used to repair circuits \cite{li2015self}, purify water \cite{pinto2020application}, and shatter blood clots \cite{cheng2014acceleration,manamanchaiyaporn2021molecular}.

All these applications rely on our ability to predict how a given matter collective reacts to driving. To be concrete here, imagine a swarm of particles being pushed around by an external field. We need to be able to predict the swarm's movements, and how those movements change as we change parameters -- to predict its collective dynamics and bifurcations. Predicting these however is hard, because of swarms' numerous degrees of freedom and nonlinear particle interactions. It takes us into the world on nonequilibrium statistical mechanics and many-body dynamical systems where standard tools and techniques fail. Take magnetic domain walls. Each one obeys the integro-differential Landau-Lifshitz-Gilbert equation, so $N$ obey a set of coupled such equations whose solution is virtually impossible. (For $N = 1,2$ walls approximations such as the $(q,\phi)$ model \cite{slonczewski1972dynamics,nasseri2018collective} have been derived, but bifurcations and scaling beyond $N>2$ is difficult \cite{nasseri2018collective,haltz2021domain}). And for magnetic colloids, because of the coupling to the host fluid, there is an extra Navier-Stokes type equation added to the mix. Vicsek-type models are sometimes used as approximations here \cite{liu2021activity,zottl2023modeling,sandor2017dynamic,reichhardt2016depinning}, but are still largely intractable; order parameters and bifurcation are often computed numerically \cite{levis2019activity,liebchen2017collective,liu2021activity,sandor2017dynamic}, leaving solvable models of driven matter scarce. 

%Unlocking the applied power of driven collectives requires an understanding of their dynamics, bifurcations, and response to forcing  -- together, these effects comprise the systems' tuning knobs. Rigorous theories of such systems are however hard to develop. Magnetic domain walls obey the formidanble integro-differential Landau-Lifshitz-Gilbert (LLG) equation; for $N = 1,2$ walls approximations such as the $(q,\phi)$ model \cite{slonczewski1972dynamics,nasseri2018collective} have been derived, but computing stabilities and scaling beyond $N>2$ is difficult \cite{nasseri2018collective,haltz2021domain}. Colloids obey hydrodynamic equations even more daunting than the LLG equation. Minimal, Vicsek-type models are sometimes used as approximations here \cite{liu2021activity,zottl2023modeling,sandor2017dynamic,reichhardt2016depinning}, but are still largely intractable; order parameters and bifurcation curves are often computed numerically \cite{levis2019activity,liebchen2017collective,liu2021activity,sandor2017dynamic}, leaving solvable models of driven matter scarce. 

This Letter helps close this research gap by introducing a model of driven matter which may be solved exactly. Our approach is to study a deliberately simplified model which hopefully captures behavior with some universality, as opposed to a detailed model specific to magnetic particles or colloids. The model's form is inspired from studies of coupled oscillators \cite{strogatz1989collective,o2022ring, yoon2022solvable,strogatz1988simple,strogatz1989predicted} which allows us to leverage new tools from that field to solve it. 

Consider $N$ particles moving in a 1D periodic domain obeying

% What we present
%This Letter presents one such model whose dynamics may be analyzed exactly. Our approach is to study a radically simplified model which hopefully captures behavior with some universality, as opposed to a detailed model specific to magnetic particles  or colloids. Consider $N$ particles moving in a 1D periodic domain obeying
\begin{align}
    \dot x_i &= E - b \sin (x_i-\alpha_i) + \frac{J}{N} \sum_j \sin (x_j - x_i) \cos (\theta_j - \theta_i), \label{eq1}\\
    \dot \theta_i &= E - b \sin (\theta_i-\beta_i) + \frac{K}{N} \sum_j \sin (\theta_j - \theta_i) \cos (x_j - x_i). \label{eq2}
\end{align}
Here $x_i,\theta_i$ are the $i$-th particle's position and phase, respectively. This phase could represent the orientation of a particle, the (in-plane) orientation of an electric or magnetic dipole, or be associated with an internal rhythm, like the chemical oscillation on the surface of an autocatalytic particle \cite{jin2023enzymatic}. Domain disorder is modeled by the $b \sin(\cdot)$ terms, which pin $x_i$ and $\theta_i$ to sites $\alpha_i, \beta_i$ (we do not consider thermal disorder), and external driving by the $E$ terms. The Kuramoto $ \sin(\cdot) \cos(\cdot)$ terms capture particle interactions. For the phases $\theta_i$, this creates synchronization which depends on the distance, for the positions $x_i$, aggregation which depends on the particles' phases, natural choices of interaction because they occur in diverse systems such as Janus particles \cite{yan2012linking} and Quincke rollers \cite{zhang2020reconfigurable}. We ignore excluded volume interactions for simplicity. %This model was inspired from earlier works on coupled oscillators \cite{strogatz1989collective,o2022ring, yoon2022solvable,strogatz1988simple,strogatz1989predicted}. 

%This model was inspired from earlier works on charge density waves \cite{strogatz1989collective, strogatz1988simple,strogatz1989predicted} and swarmaltors \cite{o2022ring,yoon2022solvable,sar2023pinning}. 

%This model was inspired from earlier works on charge density waves and swaramaltors coupled oscillators \cite{strogatz1989collective,o2022ring, yoon2022solvable,strogatz1988simple,strogatz1989predicted}. 

\begin{figure*}[t!]
\centering 
\includegraphics[width = 2\columnwidth]{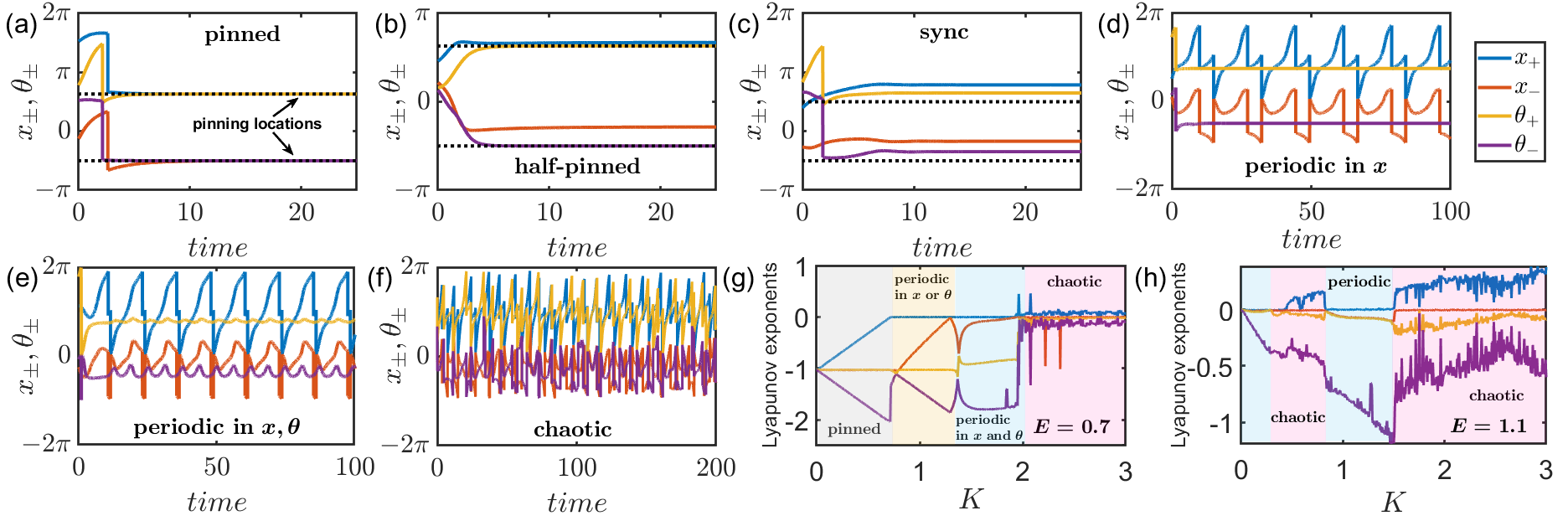}
\caption{\textbf{States and bifurcations for $J=-K$}. (a) Pinned ($E=0.4,K=0.5$), (b) half-pinned ($E=0.4,K=1.1$), (c) sync ($E=0.4,K=2.1$), (d) periodic in $x$, fixed in $\theta$ ($E=0.7,K=1.0$), (e) periodic in both $x$ and $\theta$ ($E=0.7,K=1.8$), (f) chaotic ($E=1.1,K=1.0$). Lyapunov exponents highlighting the bifurcations for two different values of $E$. (g) $E=0.7$, and (h) $E=1.1$. The system is integrated with $(dt,T) = (0.01,100)$ using an RK4 method.}
\label{states}
\end{figure*}

\textit{Small $N$ regime}. We explore the little $N$ limit with a case study of $N = 2$ particles. The physical system we have in mind here is two magnetic domain walls moving on circular race track memory \cite{zhang2018manipulation},  similar to recent experiments \cite{hrabec2018velocity}, \footnote{There, however, the spatial domain was a straight line; here it is periodic}. For simplicity, we study symmetric pinning sites $(\alpha_1, \alpha_2) = (\beta_1, \beta_2) =  (0,\pi)$ \footnote{Which may be appropriate for periodic substrates \cite{reichhardt2016depinning,hoffmann2000periodic}}. Setting $b=1$ without loss of generality and moving to coordinates $(x_{\pm}, \theta_{\pm}) = ((x_1 \pm x_2)/2, (\theta_1 \pm \theta_2$)/2) yields 
\begin{align}
    & \dot{x}_+ = E -  \cos x_+ \sin x_- \label{xp}, \\
    & \dot{x}_- = - \sin x_+ \cos x_-  - \frac{J}{2} \sin 2 x_- \cos 2 \theta_- \label{xm},   \\
    & \dot{\theta}_+ = E - \cos \theta_+ \sin \theta_- \label{thetap},  \\
    & \dot{\theta}_- =- \sin \theta_+ \cos x_-  - \frac{K}{2} \sin 2 \theta_- \cos 2 x_-  \label{thetam}.
\end{align}
%\begin{figure}[t!]
The behavior of the model divides into two cases depending on the relative sign of $J$ and $K$. We present the opposite sign case here (it contains the most relevant physics) and the same-sign case in the supplementary material (SM). We also set $J = -K$ for ease, since the magnitude of $J$ does not change the overall phenomena (see SM).

First we develop some intuition for our system by visualizing its dynamics. Imagine the particles as moving dots in the $(x,\theta)$ plane with periodic boundary conditions \footnote{Equivalently, the torus}. Limit case dynamics are easy to picture. When the driving dominates $E \gg b,K$, the particles will be swept around the plane in uniform rotations. When the pinning is large $b \gg E,K$, they will freeze into their pinning sites $x_i = \theta_i = \alpha_i$. And when the coupling wins out $K \gg b,E$, they will unstick from $\alpha_i$ and lock into synchrony \footnote{since $J=-K$, it won't be straightforward synchrony $x_1=x_2, \theta_1 =\theta_2$, which we expect for $J=K$, but some type of anti-sync}. But what happens when the three effects have comparable strength $b \approx K \approx E$? And how do the various states arise and disappear as the parameters change?

\begin{figure}[t!]
\centering 
\includegraphics[width = 0.9 \columnwidth]{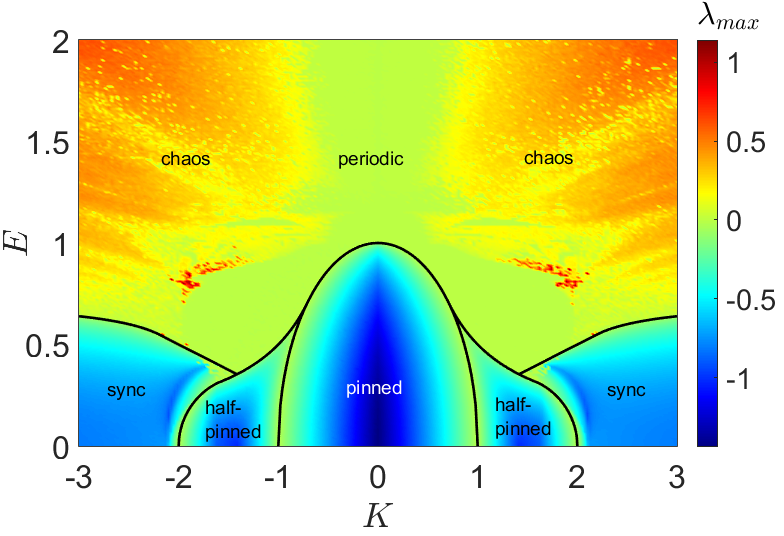}
\caption{\textbf{Bifurcation diagram for $N=2$ particles}. Black lines denote theoretical predictions, colors signify the largest Lyapunov exponent. In the periodic / chaotic parameter regions, bistability was sometimes observed. }
\label{bif-diagram-J-minus-K}
\end{figure}

To answer these questions, we ran numerical experiments. Figure~\ref{states} shows our results, but before we talk through them, you should look at the bifurcation diagram in Fig.~\ref{bif-diagram-J-minus-K} which encapsulates all the results at once. Having this birds eye view of the systems dynamics in mind as you read will help you follow the story. Supplementary Movie 1 also presents a live demo of our experiments which is also useful to watch at this point.

To begin, we realized the pinned state -- a natural `ground state' to perturb around -- by turning off the driving $E=0$ and setting the phase coupling $K$ small. Fig.~\ref{states}(a) shows time series of the coordinates $x_{\pm}, \theta_{\pm}$ relaxing from random IC to the pinning sites, depicted as dotted lines (in the $\pm$ coordinates, the sites are $\alpha_{\pm} = \pm \pi/2$). Then we gradually increased $K$, expecting the particles to synchronize, in the sense of minimizing their space differences $x_-$ and phase differences $\theta_-$. (Note however that since driving is turned off, we don't expect the particles to oscillate here, we expect them to just shift to a new fixed point). We find two transitions. First, a half pinned state emerges, where $\theta_{\pm}$ stay pinned, but the phases $x_{\pm}$ are locked into a new sync fixed point (Fig.~\ref{states}(b)). Notice the $\theta_{\pm}$ line settles onto the pinning sites (dotted line), but $x_{\pm}$ does not. In the second transition, the remaining coordinates depin and a full sync state is realized where all four coordinate lie off the pinning sites (Fig.~\ref{states}(c)).

\begin{figure*}[t!]
\centering 
\includegraphics[width = 2\columnwidth]{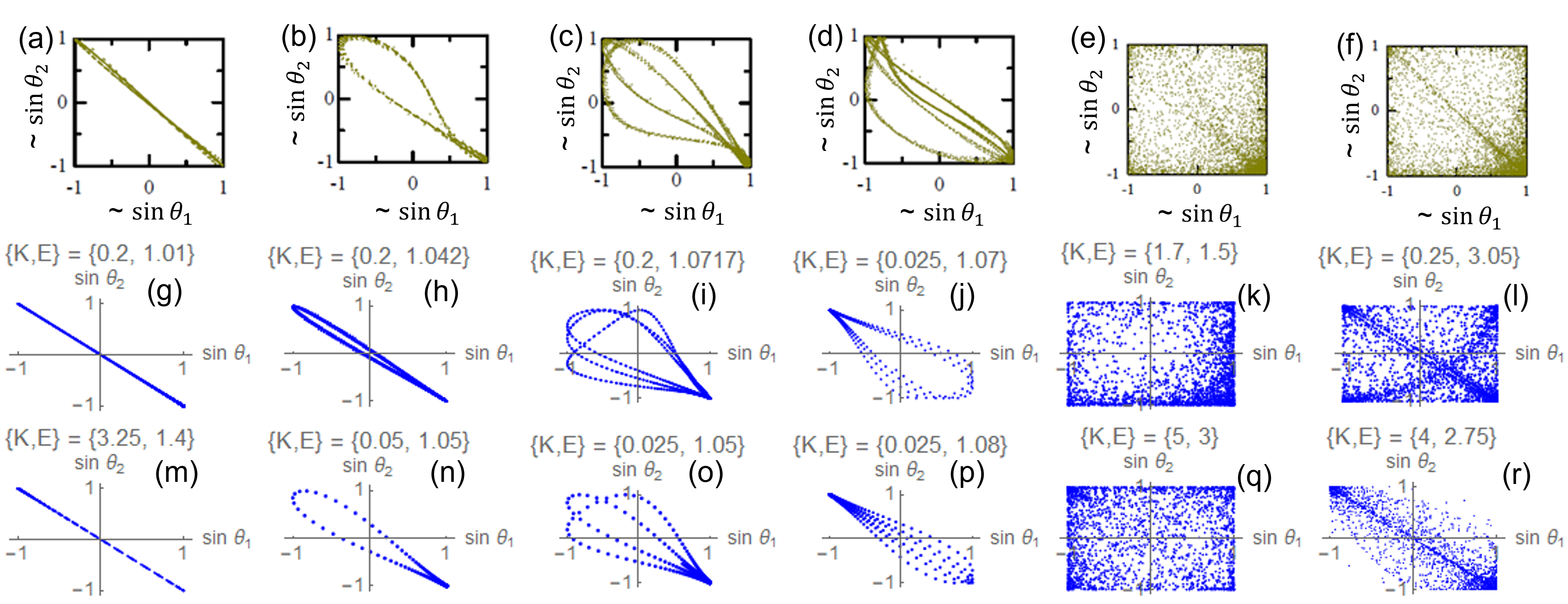}
\caption{\textbf{Match to micromagnetic simulations of two domain walls}. Top row (a)-(f) high resolution micromagnetic simulations from \cite{hrabec2018velocity}. Middle row (g)-(l), $J=-K$ and bottom row (m)-(r), $J=K$ model Eqs.\ \eqref{eq1}-\eqref{eq2}. Numerical parameters: $(dt,T) = (0.25,200)$. Initial $(x_i, \theta_i)$ were drawn uniformly at random from $[0,2\pi]$. Multistability was sometimes observed. Reprinted Supplementary Figs. 7,8 with permission from [Ale{\v{s}} Hrabec et al., Physical Review Letters, 120, 227204 (2018)]. Copyright (2023) by the American Physical Society.}
\label{match}
\end{figure*}

%The transitions between the three states is also depicted in the bifurcation diagram Figure~\ref{bif-diagram-j=-k}.

% improve this. not clear. Fix figures.
%Answers to these questions turn out to be quite delicate. Numerics show that for low driving and increasing $K$, the pinned state bifurcates twice: first into a half-pinned state where either $\theta_i$ or $x_i$ unlock and the other stays pinned, and then into a sync state where both $x_i, \theta_i$ unlock. Figs.~\ref{states}(a)-(c) show the relaxation of $x_{\pm}(t), \theta_{\pm}(t)$ to these static states, with the pinning sites denoted by dotted lines \footnote{When $E>0$, the pinning sites deviate from $(\alpha_0, \alpha_1) = (0,\pi)$ slightly which means in sum/difference coordinates $y_{-} = (\alpha_1 - \alpha_2) / 2 = (0-\pi)/2 = -\pi/2$, where $y_- = x_-, \theta_-$, but $y_+ = (\alpha_1 + \alpha_2) / 2 = -\cos^{-1}(E)$. See analysis section.}. The bifurcation diagram in Fig.~\ref{bif-diagram-J-minus-K} depicts the states' transitions; for $E \lesssim 0.25$, the stability regions for the states are adjacent to each other.

How much driving $E>0$ can the three states withstand before $x_{\pm}, \theta_{\pm}$ unlock and start moving? Look at the bifurcation diagram in Fig.~\ref{bif-diagram-J-minus-K}. For $E \lesssim 0.25$, the pinned $\rightarrow$ half-pinned $\rightarrow$ sync transition persists, but for larger driving $E \approx 0.5$, some periodic states (green region) arise between the half-pinned and sync states. Here, either two of the four coordinates $(x_\pm, \theta_{\pm})$ oscillate, the others remaining fixed (Fig.~\ref{states}(d)), or all four coordinates oscillate (Fig.~\ref{states}(e)). For larger driving still $E \approx 0.8$, the pinned state morphs directly into periodicity, which in turn undergoes an intermittency transition to chaos (orangish region) depicted in Fig.~\ref{states}(f).  Finally, for $E > 1$, the static states vanish and the dynamic ones become reentrant: the system flip flops between chaotic and periodic motion, then settles into chaos (bistability between the two states was sometimes observed). To confirm the motion was chaotic, we computed a heat map of Lyapunov exponents $\lambda_{max}$ in the $(K,E)$ plane and saw $\lambda_{max}>0$ where expected. We also compute power spectra which indicate the chaos transition is of the intermittent type (Fig. S2). For convenience, we also plot the Lyapunov exponents $\lambda(K)$ to show the bifurcation sequence at $E=0.7$ and $E=1.1$ in Fig.~\ref{states}(g)-(h). 

Let's take stock of our findings. We find three static states, a family of periodic states, chaos, and various interstate transitions. To check these states' robustness, we run simulations for uneven coupling $J=-cK$ for $c\neq1$ and asymmetric pinning $(\alpha_1, \alpha_2) = (0, a \pi)$ for $0\leq a \leq 1$ \footnote{We can set $\alpha_1 = 0$ without loss of generality} and find the same physics (qualitatively identical bifurcation diagrams; see Figs. S5,S6,S7). A surprise for asymmetric pinning (Fig. S6) is that the bifurcations of the static states get richer; the half-pinned and sync states become reentrant. Most of the collective states also appear for same-sign coupling $J=K$, albeit with a different bifurcation structure (Fig. S1).

%Let's take stock of our findings. We found three static states, a family of periodic states, chaos, and various interstate transitions. To check these states' robustness, we ran simulations for $N=2,3,4,5$ particles, uneven coupling $J=-cK$ for $c\neq1$, and asymmetric pinning $(\alpha_1, \alpha_2) = (0, a \pi)$ for $0\leq a \leq 1$ \footnote{We can set $\alpha_1 = 0$ without loss of generality} and found the same physics (qualitatively identical bifurcation diagrams; see Figs. S5,S6,S7,S8).  A surprise for asymmetric pinning (Fig.S6) was that the bifurcations of the static states get richer -- notice the half-pinned and sync states become reentrant. Most of the collective states also appeared for same-sign coupling $J=K$, albeit with a different bifurcation structure (SM, Fig. S10).

% Improve description of experiment / simulation. State why we can think use one theta_i. Maybe refer to an SM picture. 
% Our hope was that our minimal model discoverd states are are universal. They do. Mag domain walls. Force-velocity curves.
% Also extend the discussion of the superconductor vortices. 
% Relegate the J=K match to SM, or else 
% Also: be sure to remind the reader that the main
%case is J = -K. And that we plot the J = + K to show the states are robust.

Our hope was that these states capture real world behaviour. They do. Figure~\ref{match} shows they mimic the behavior found in a recent study \cite{hrabec2018velocity} of a pair magnetic domain walls. Briefly, their setup is this. Each wall is free to move in the $x$ direction, and one wall is slightly higher than the other in the $y$ direction, so that they do not collide. Then each wall $i$ may be characterized by a single spatial degree of freedom $x_i \in \mathbb{R}^1$ and also a phase $\theta_i \in \mathbb{S}^1$ corresponding to the effective magnetic dipole vector of the wall; thus the walls fall into our model class \footnote{see \cite{hrabec2018velocity} for more details on the phase $\theta_i$}. The walls begin pinned at fixed points $(x_i^*, \theta_i^*)$. Then an external magnetic field is turned on, which induces their positions $x_i$ and $\theta_i$ to unlock and interact. The top row of Fig.~\ref{match}, taken from \cite{hrabec2018velocity}, shows the resulting dynamics for different parameters with scatter plots of the dipole vectors for each wall $(m_1, m_2)$. These $m_i$ relate to the phase via $ m_i \propto \sin \theta_i$, as indicated by the axes labels.  Notice the walls settle into simple periodic behavior or more complex dynamics represented by Lissajous curves and point clouds. The middle column shows our model reproduces these states for the $J=-K$ case study we have presented in the main text. The bottom row shows the same-sign $J=K$ case, whose analysis, recall, we present in the SM, can also reproduce the states.

Beyond magnetic domain walls, our model can also reproduce realistic force-velocity curves $\bar{v}(E)$ which are standard ways to analyze pinned-driven systems (here $\bar{v}$ is the average system velocity and $E$ is the driving force). Fig. S9 shows the $\bar{v}(E)$ of our model imitates that of pinned superconductor vortices \cite{reichhardt2016depinning}, for example.

Now we turn to analysis. We derive fixed point expressions for $x_{\pm}, \theta_{\pm}$ in the static states and derive their bifurcation curves drawn as black lines in Fig.~\ref{bif-diagram-J-minus-K}. We find the fixed points by setting the RHS of Eqs.~\eqref{xp}-\eqref{thetap} to zero. Then we eliminate $(x_-, \theta_-)$ using Eqs.~\eqref{xp}, \eqref{thetap} and sub the result into Eqs.~\eqref{xm},~\eqref{thetam}:
%Now we explain our numerical findings analytically. The first move is to eliminate $(x_-, \theta_-)$ using Eqs.~\eqref{xp}, \eqref{thetap} and sub the result into Eqs.~\eqref{xm}, \eqref{thetam}:
\begin{align}
&\sqrt{1-E^2 \sec ^2 x_+} \Big( \frac{E K \left(2  E^2 \sec ^2 \theta _+ -1\right)}{\cos x_+}-\sin x_+ \Big)  = 0, \label{e1}  \\
&\sqrt{1- E^2 \sec ^2 \theta _+} \Big( \frac{E K \left(2 E^2 \sec^2 x_+-1\right)}{\cos \theta_+}-\sin \theta _+ \Big) = 0. \label{e2}
\end{align}
The systems has form $AB = CD = 0$, implying four sets of fixed points. When $(A,C) = (0,0)$ we get the pinned state. When the $(A,D) = (0,0)$ or $(B,C) = (0,0)$ we get the half pinned state (in the one $x_i$ stays pinned, $\theta_i$ syncs, in the other the reverse), and when $(B,D) = (0,0)$ we get the sync state. The analysis of each state is complex, but the strategy is simple: find the fixed points, and then their stability by linearization. So we present just a sketch of the analysis for the pinned state here, and put the rest in the SM.

The condition $(A,C) = (0,0)$ implies $\sqrt{1-E^2 \sec^2 x_+} = \sqrt{1- E^2 \sec^2 \theta_+} = 0$. Solving these yields a family of 16 fixed points, $(x_+, x_-, \theta_+, \theta_-) = (\pm \cos^{-1} (\pm E), \pm \pi/2, \pm \cos^{-1} (\pm E), \pm \pi/2)$, four of which are stable, reflecting a single macrostate. Linearization reveals a zero eigenvalue at $E_c = \sqrt{1 - K^2}$. Turning the same crank for the other states reveals two-piece boundaries: for the half-pinned state, $E_c(K) = 1/2K$ and $E_c(K) = \sqrt{-4+ 5K^2-K^4}/3K$, intersecting at $K^* = \sqrt{5/2}$, and for the sync state, $E_c = K/4$, $E_{c} = 1/2 \sqrt{(K (\sqrt{K^2-4}+K)-1)/K^2}$ which meet at $K^* = \sqrt{2 (1+\sqrt{2})}$. The expressions for $x_{\pm,},\theta_{\pm}$ for these states are too complex to display here. This completes our analysis of the small $N$ regime.

\begin{figure}[t!]
\centering
\includegraphics[width = \columnwidth]{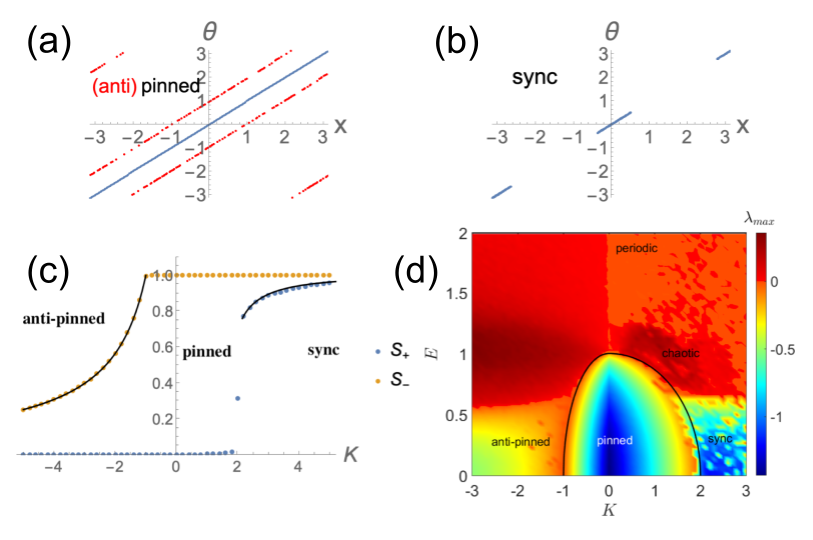}
\caption{\textbf{Large $N$ limit for $J=K$}. (a) Pinned state (blue, $K=1$) and anti-pinned state (red, $K=-2$) plotted on same graph to save space for $E=0$ and $N=200$ particles. (b) Sync state for $(K,E)=(3,0)$. (c) Order parameters for $E = 0$. (d) Bifurcation diagram in $(K,E)$ space. Black curves show theoretical predictions. Colors denote Lyapunov exponents, computed using $N=10$ particles, which well approximated the $N\gg1$ limit (simulations for larger $N$ were prohibitive). }
\label{largeN}
\end{figure}
\textit{Large N regime}. How do our results change as we scale up the population size $N \gg 1$? A surprise is the $J=-K$ case we have thus far considered gets simpler. The half-pinned and sync states disappear, leaving just the pinned and unsteady states (Fig. S8). We used linearly spaced pinning $\alpha_i = 2 \pi i / N$ here to facilitate analysis (randomly chosen $\alpha_i, \beta_i$ produce similar results \cite{sar2023pinning}).

Dynamics for same sign coupling $J=K$, in contrast, get richer. Fig.~\ref{largeN}(a) shows the pinned state (blue dots) persists, but now a new anti-pinned state (red dots) arises where neighboring particles are shifted by an amount $x_{i+1} - x_{i} = \theta_{i+1} - \theta_i = \Delta(K)$. Fig.~\ref{largeN}(b) shows the large $N$ analogue of the sync state with particles bunching into two sync clusters. New periodic behavior is observed, along with quasiperiodicity, and the route to chaos is now via period doubling. Supplementary Movies 2,3 depict the evolution of all the states and Fig. S4 studies the transition to chaos.

We analyzed the $N \rightarrow \infty$ states in a previous work \cite{sar2023pinning}. The critical coupling for the anti-pinned state is $E_c =  \sqrt{1 - K^2}$, found using a self-consistency analysis, and for the pinned state is $E_c =  \sqrt{1 - K^2/4}$, found using a variational argument \footnote{Note, in \cite{sar2023pinning} we called pinned, anti-pinned states the phase wave, and split phase wave. These are tools from coupled oscillator theory we advertised in the introduction.}. These are the sides of the lop-sided bell in the bifurcation diagram Fig.~\ref{largeN}(d). We also derived expressions for the order parameters
\begin{align}
    S_{\pm} e^{i \Phi_{\pm}} = \frac{1}{N} \sum_j e^{i(x_j\pm\theta_j)}  \label{S_def}.
\end{align}
In the split phase wave,
\begin{align}
    S_- &= -\frac{1}{3} + \frac{K^4 + (\Gamma_1 + 3 \sqrt{3} \sqrt{ \Gamma_2} )^{2/3}}{3 K^2 (\Gamma_1 + 3 \sqrt{3} \sqrt{ \Gamma_2})^{1/3} } , \label{smE} 
\end{align}
where
\begin{align}
    \Gamma_1 & := -27 \left(E^2-1\right) K^4-K^6 , \\
    \Gamma_2 & := \left(E^2-1\right) K^8 \left(27 E^2+2 K^2-27\right) .
\end{align}
In the sync state, we derived a pair of self-consistency equations for $S_+$
\begin{align}
    -& 2 \sin \Big( \frac{\xi}{2} - \alpha \Big) + K S_+ \sin (\Phi_+ - \xi)  = 0, \label{s1} \\
    & S_+ e^{\Tilde{i} \Phi_+} = \frac{1}{2\pi}\int_{0}^{2\pi} e^{\Tilde{i} \xi(\alpha)} d \alpha , \label{s2} 
\end{align}
where $\xi_i := x_i -\theta_i$ which we solved numerically \footnote{We must solve Eq.~\eqref{s1} for the fixed points $\xi^*(\alpha)$ then plug them into Eq.~\eqref{s2} to find $S_+$. First we set $\Phi_+ = 0$ without loss of generality. Then by applying various trig identities to Eq.~\eqref{s1} we arrive 4-th order polynomial in $\cos \xi^*$ and plug the roots into Eq.~\eqref{s2}, which when $\Phi_+ = 0$ reads $S_+ = \int \cos(\xi^*(\alpha)) d \alpha$. Then we computed the integral over $\alpha$ numerically.}. Figure~\ref{largeN}(c) shows these $S_{\pm}(K)$ match simulations and distinguish between the static states; in the pinned state $(S_+, S_-) = (1,0)$ trivially by subbing $x_i = \theta_i$ into the definition for $S_{\pm}$ Eq.~\eqref{S_def} . There is also a small region of hysteresis (not visible in the graph) between the pinned and sync states \cite{sar2023pinning}. This completes our analysis.

%Exact analyses of driven active matter are rare. Their numerous degrees of freedom and non-equilibrium nature make a mathematical studies difficult. This Letter helps shed analytic light on this class of problem by introducing a toy model tractable in both the low and large $N$ limit. Surprisingly, the intermediary $N$ regime is essentially. 

% Main point: we may have found the "Ising model" of pinned 1D systems. Our main finding is this: we have found a solable model. 
% Its so simple, while at the same time mimicks. Could it be the Ising model for 1D pinned system? Future work could help determine 
% This claim by trying to derive the 1D model from the. Start with the rigourous LLG equations, they use phase reduction techniques, and 
% see if our simple model pops out. It would be fantastic if that happened. It would. 

The reaction of particulate matter to external driving is crucial for applications, yet difficult to understand theoretically. This Letter sheds light on this class of dynamics with a toy model tractable in both the low and large $N$ limit -- moreover, the intermediary $N$ regime is surprisingly well approximated by the $N \rightarrow \infty$ model; Figures S8, S10 show as few as $N \approx 4$ particles give the same physics. The model also captures the behavior of real world systems such as magnetic domain walls (Fig. 3), Japanese tree frogs \cite{aihara2009modeling} and Janus matchsticks \cite{chaudhary2014reconfigurable} (both realize the anti-pinned state Fig.~\ref{largeN}(a)), and superconductor vortices ($\bar{v}(E)$ curve in Fig. S9). We also suspect its chaos may be connected to the active turbulence of biological microswimmers \cite{das2020transition}, since the swimmers contain the same basics physics as the model: driving and emergence in environments with pinning.

Given this balance between solvability and realism, we wonder if the model `could be the Kuramoto model' for this class of driven matter, by which we mean the simplest, representative model in a universality class. Future work could explore this conjecture by deriving our model from a physically rigorous model. Start with say the Landau-Lifshitz-Gilbert equations for magnetic particles, exploit a small quantity like a weak coupling limit or a separations of time scales using a perturbative method \cite{kuramoto}, and see if our model or something close to it pops out. This was the way the Kuramoto model itself was derived -- starting with a general reaction diffusion equation and simplifying using phase reduction methods -- and is the source so to speak of its universality \cite{kuramoto}.

%The reaction of particulate matter to external driving is crucial for applications, yet difficult to understand theoretically because of the systems' nonlinearities and numerous degrees of freedom. This Letter sheds light on this dynamics class with a toy model tractable in both the low and large $N$ limit -- moreover, the intermediary $N$ regime is surprisingly well approximated by the $N \rightarrow \infty$ model; Figures S4-5 show as few as $N=5$ particles give the same physics. The model also captures the behavior of real world systems such as magnetic domain walls (Fig. 3), Japanese tree frogs \cite{aihara2009modeling} (anti-pinned state), Janus matchsticks \cite{chaudhary2014reconfigurable}  (anti-pinned state), and superconductor vortices ($\bar{v}(E)$ curve in Fig. S11). We also suspect its chaos may be connected to the active turbulence of biological microswimmers \cite{das2020transition}, since the swimmers contain the same basics physics as the model: driving and emergence in environments with pinning.

Our model could guide experimental work on systems of magnetic domains walls and other particles \cite{hrabec2018velocity,ledesma2023magnetized}. Chaos has a niche application in such systems, it can be exploited for hardware security \cite{ghosh2016spintronics}), but to our knowledge has not yet been reported in multiparticle experiments \footnote{Chaos has been observed in single particle systems \cite{chen2022chaotic,shen2020current}}. Our model predicts it occurs for all $N>1$ and gives parameter regimes where it is likely to arise. The model or a close variant may also be useful in studies of charge density waves which couple to lattice vibrations \cite{hedayat2019excitonic}. Eq.~\eqref{eq2} for $\dot{\theta}_i$ has already been used to model the phase of the charge wave \cite{strogatz1988simple,strogatz1989predicted,strogatz1989collective}; the novelty would be Eq.~\eqref{eq1} for $\dot{x_i}$ which could represent the displacements of the lattice atoms from their equilibrium positions. Such displacements form sinusoidal patterns \cite{hedayat2019excitonic}, encapsulated by the $b \sin(x_i - \beta_i)$ term, which competes with the tendency to `synchronize' at equilibrium $ x_i= x_j = 0$, as per the $K \sin(x_j - x_i)$ term.

%Future work could try to connect our model to the magnetic LLG equation using phase reduction or other perturbative methods \cite{kuramoto}. Generalizing the model to motion in 2D, or relaxing its idealized mean field coupling, would also be interesting. 

%%%%%%%%%%%%%%%%%%%%%%%%%%%%%%%%%%%%%%%%%%%%%%%%%%%%%%%%%%%%%%%%%%%%%%%%%%%%%%%%%%%

%%%%%%%%%%%%%%%%%%%%%%%%%%%%%
\textbf{Reproducibility}. Code used for simulations and analytic calculations available at github \footnote{https://github.com/Khev/swarmalators/tree/master/1D/on-ring/random-pinning}.

%%%%%%%%%%%%%%%%%%%%%%%%%%%%%%%%%%%%%%%%%%%%%%%%%%%%%%%
    
\bibliographystyle{apsrev}

%\bibliography{ref.bib}

\end{document}